\documentclass[]{aa}
\usepackage{graphicx}
\usepackage{txfonts}
\usepackage{ulem}
\def\mgls{1408}  %
\def\mirm{-25}  %
\def\pfe{0.18}  %
\def\sfe{0.24}  %
\def\tse{1.7}  %
\def\hane{0.8}  %
\def\llmf{20}  %
\def\radms{\hbox{rad$\,$m$^{-2}$ }}  
\def\radm{\hbox{rad$\,$m$^{-2}$}}  
\def\lamAa{\hbox{$\lambda\,21\,\mbox{cm}$}}  
\def\lamA{\hbox{$\lambda\,21\,\mbox{cm}$ }}  
\def\lamB{\hbox{$\lambda\,18\,\mbox{cm}$ }}

\begin{document}

\title{Faraday screens associated with \\ local molecular clouds\thanks{Based on 
observations with the Effelsberg 100-m telescope
operated by the Max-Planck-Institut f\"ur Radioastronomie (MPIfR), Bonn, Germany}}

\author{M. Wolleben, and W. Reich}

\offprints{M. Wolleben}

\institute{Max-Planck-Institut f\"{u}r Radioastronomie, Auf dem H\"{u}gel
69, 53121 Bonn, Germany\\
 email: wolleben@mpifr-bonn.mpg.de, wreich@mpifr-bonn.mpg.de }

\date{Received ; accepted }

\abstract{Polarization observations at \lamA and \lamB towards the 
local Taurus molecular cloud complex were
made with the Effelsberg 100-m telescope. Highly structured, frequency-dependent polarized emission
features were detected. We discuss polarization minima with excessive rotation measures located
at the boundaries of molecular clouds. These minima get less pronounced at the higher frequencies. 
The multi--frequency polarization data have been successfully modeled by considering magneto--ionic Faraday 
screens at the surface of the molecular clouds. Faraday rotated background emission adds to foreground 
emission towards these screens in a different way than in its surroundings.  The physical size of the
Faraday screens is of the order of $2$~pc for $140$~pc distance to the Taurus clouds. Intrinsic 
rotation measures between about $-18$~\radms to $-30$~\radms are required to model the observations. 
Depolarization of the background emission is quite small (compatible with zero), 
indicating a regular magnetic field structure with little turbulence within the Faraday screens. With 
observational constraints for the thermal electron density from H$\alpha$ observations
of less than $\hane$~cm$^{-3}$ we conclude that the regular magnetic field strength along the line of sight 
exceeds $\llmf~\mu$G, to
account for the observed rotation measure. We discuss some possibilities for the origin of such strong 
and well ordered
magnetic  fields. The modeling also predicts a large-scale, regularly polarized emission in the foreground of the Taurus
clouds which is of the order of $\pfe$~K at \lamAa. This in turn constrains the observed synchrotron emission in
total intensity within $140$~pc of the Taurus clouds. A lower limit of about $\sfe$~K, or 
about $\tse$~K/kpc, is required for a perfectly 
ordered magnetic field with intrinsic ($\sim75\%$) percentage polarization. 
Since this is rather unlikely to be the case, the fraction of foreground synchrotron emission is even larger. 
This amount of synchrotron emission is clearly excessive when compared to previous
estimates of the local synchrotron emissivity. 
\keywords {linear polarization -- Faraday rotation -- local synchrotron emission -- magnetic field -- ISM -- Taurus molecular clouds}}

\titlerunning{Faraday screens near molecular clouds}
\maketitle

\section{Introduction}

For large areas of the sky the polarization properties of Galactic radio emission are not well known in
detail. Most information came from early large--scale surveys of the
northern sky carried out at Jodrell Bank and Cambridge with $1.4$~GHz being the highest frequency 
used (Bingham \cite{bing}). The angular resolution of all these surveys was low. 
Later the Dwingeloo--surveys were carried out in the frequency range from $408$~MHz up to $1415$~MHz
(Brouw \& Spoelstra \cite{brouw}). They have angular resolutions of $0\fdg6$ at 1411~MHz or
less at the lower frequencies and are absolutely calibrated, but severely under-sampled. 
Details of the polarization structures require fully sampled measurements at arc-minute angular resolution. 
A number of observations were made during the last decade using
synthesis arrays as well as single-dish telescopes. 
Polarization observations by Junkes et al. (\cite{junk}), Wieringa et al. (\cite{wie}), 
Duncan et al. (\cite{dunc97}, \cite{dunc99}),
Gray et al. (\cite{gray1}, \cite{gray2}),  Uyan{\i}ker et al. (\cite{buII}), Haverkorn et al. (\cite{haver}),  
Gaensler et al. (\cite{gaens}) revealed highly varying polarized structures,
which are not correlated to structure variations seen in total intensity maps. 
These findings called for a systematic approach and new polarization surveys
were initiated.

Here we report on follow-up observations at \lamB of the Effelsberg \lamA \char`\"{}Medium 
Galactic Latitude Survey\char`\"{} (EMLS, Reich et al., in preparation). This survey covers the 
northern Galactic plane in the range of $\pm 20^{\circ }$. The first EMLS maps (Uyan{\i}ker 
et al. \cite{buII}) show an unexpected richness in structure of the polarized emission, which
indicates significant variations in the properties of the magneto-ionic
medium. Since the polarization structures are not associated with total
intensity enhancements or depressions, the interpretation of Faraday
modulation of an intrinsically polarized background seems the most
likely explanation for the observations.

The physical parameters of the Faraday rotating regions, the Faraday screens, are of particular 
interest as they reflect local conditions of the magneto-ionic interstellar medium, which deviate
from the average parameters derived from the analysis of rotation measure ($RM$)
from pulsars or extragalactic sources. The average density of thermal electrons in the Galactic 
plane is about $0.03$~cm$^{-3}$ and the average of the regular magnetic field along the line of 
sight is about $1$ to $2~\mu$G (Taylor \& Cordes \cite{tay}; Comez et al. \cite{com}; 
Cordes \& Lazio \cite{cord}). 
However, local bubbles with $RM$s exceeding $100$~\radms are known
to exist in the Galaxy (e.g. Gray et al. \cite{gray1}; Gaensler et al. \cite{gaens}; 
Uyan{\i}ker et al. \cite{buIII}).
Such large $RM$ values need kpc-size structures to originate within the average magneto-ionic
medium and strongly suggest significant local enhancements of the magnetic field
strength and/or the thermal electron density. The total
magnetic field strength was estimated to be about $5$ to $6~\mu$G in the solar vicinity 
(Strong et al. \cite{strong}).
Even in the unlikely case that the regular component along the line of sight locally has
that value, an excessive size of the Faraday screens of
some hundred parsec is required.  

The key problem is the unknown distance of the Faraday screens and
this makes any estimate of their physical parameters rather uncertain. Distance
information can be obtained by spectral line observations of either
\ion{H}{i}, \element[][]{CO} or recombination lines of neutral hydrogen clouds, molecular
clouds or \ion{H}{ii} regions associated with Faraday screens. The Taurus--Auriga
region seems to be quite suited for such an investigation. The large molecular and dust 
cloud complexes are at a distance of about $140$~pc, and it is therefore
possible to resolve structures on pc--scales with arc-min angular resolution. In addition 
the Taurus-Auriga complex
is located at medium latitudes well below the Galactic Plane, so that confusion
with the intense emission from the Galactic disk is not important.
The maps from the EMLS revealed a number of polarized enhancements or
depressions apparently related to molecular material. In order to derive
physical
properties of the associated Faraday screens we  have complemented the
\lamA survey data of the Taurus--Auriga region by \lamB
observations to derive the depolarization properties and the rotation measures.
We modeled the polarization data in order to constrain the physical
parameters of the Faraday screens by taking into account foreground and
background emission. 

\section{Observations and data reduction}

\begin{figure*}
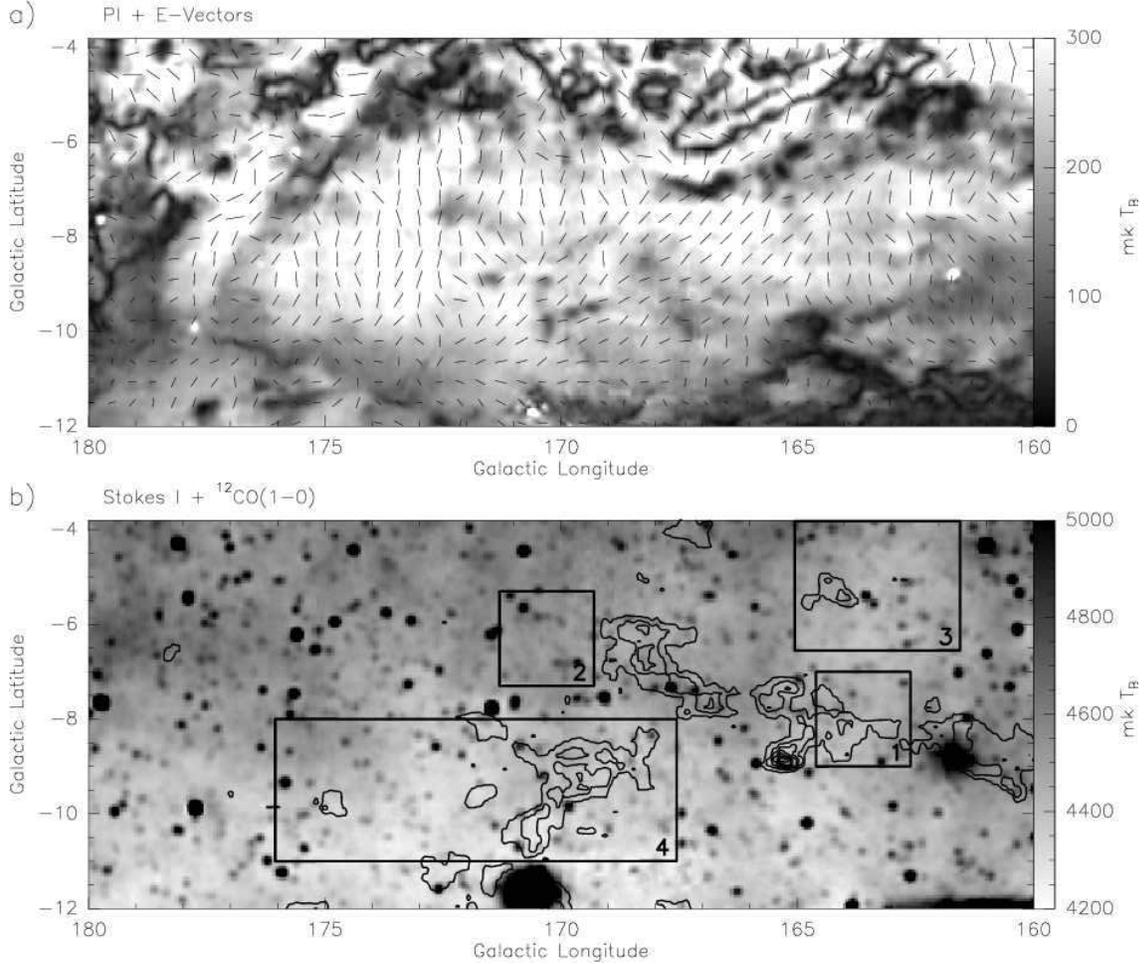

\centering
  \includegraphics[height=15cm, angle=270]{0561fig1.jeps}
  \includegraphics[height=15cm, angle=270]{0561fig2.jeps}
  \caption{{\bf a)} Section from the \lamA \char`\"{}Medium Galactic Latitude
Survey\char`\"{} including absolute calibration showing polarized
intensities with polarization vectors
overlaid in E--field direction. {\bf b)} the same for \lamA total intensities.
These include the $2.7$~K cosmic microwave background.
Overlaid contours show integrated
\element[][]{CO} emission from the survey by Dame et al. (\cite{dame}). The four fields observed at
\lamB are marked and labeled. }
\label{bildchen1}
\end{figure*}

The \lamA observations for the EMLS were recently completed and are presently in final reduction stage.
The data covering the Taurus-Auriga region extending below the Galactic
plane from $l=150^{\circ }$ to $l=190^{\circ }$ were extracted and reduced
by Wolleben (\cite{woll}) following the survey reduction procedures. In brief, total intensities 
and linear polarization were measured simultaneously with sensitivities of $15$~mK for Stokes~$I$ and
$8$~mk for Stokes~$U$ and $Q$ at 
an angular resolution of $9\farcm35$. The observing, reduction and calibration methods
of the survey and selected maps from different regions along the Galactic plane were 
already published
by Uyan{\i}ker et al. (\cite{buI}, \cite{buII}). The absolute calibration of
the total intensity data was done in the usual way by adding the missing large-scale 
structure from the absolutely calibrated Stockert \lamA
survey (Reich \& Reich \cite{rei86}). The absolute calibration of the Effelsberg polarization
survey data relied so far on Dwingeloo polarization data (Brouw \& Spoelstra \cite{brouw}). 
However, the Dwingeloo survey is rather incomplete in coverage and largely under-sampled, so only a few measurements were available for the Taurus--Auriga region. We therefore made use
of the new polarization survey carried out with the DRAO 26-m telescope at 1.41~GHz, which 
consists of fixed declination scans being fully sampled along right-ascension. They are separated
in declination by 1$^{\circ}$ to 2$^{\circ}$. A brief description of the survey methods and preliminary
results was already given by Wolleben et al. (\cite{woll1}). The DRAO data for Stokes $U$ and $Q$ were
compared with the corresponding Effelsberg maps both being convolved with a 6$^{\circ }$-wide Gaussian
and the differences were added to the original Effelsberg $U$ and $Q$ maps, respectively.

The \lamA Taurus--Auriga map 
in total intensity and polarization was discussed in detail by Wolleben (\cite{woll}) with large-scale
polarization correction using Dwingeloo data. 
A section of this map with the new large-scale DRAO polarization data as described above is displayed 
in Figs.~\ref{bildchen1}a and b showing polarized and total intensities.
Total intensities were overlayed with the
distribution of integrated \element[][]{CO} emission as observed by Dame et al. (\cite{dame}) 
tracing numerous molecular gas clouds in this region.

Follow--up observations of selected fields, where significant polarization
structure was seen, were done using the Effelsberg 100-m telescope at
two frequencies around \lamB ($1660$~MHz and $1713$~MHz) in September 2001. The same receiver 
as for the \lamA EMLS was used, but a different HF-filter was selected to
suppress interference 
out of the \lamB band. The observations were done simultaneously for
both frequencies by utilizing two identical IF-polarimeters. Both have an effective bandwidth of $14$~MHz. 
The fields were mapped two times in orthogonal scanning directions along Galactic latitude and
Galactic longitude, in a similar way to the
\lamA EMLS observations. The
angular resolution was $7\farcm87$ at $1660$~MHz and $7\farcm70$ at $1713$~MHz,
respectively.  The maps were fully sampled at
$3\arcmin$. \object{3C286} served as the main calibration source with a flux density of
$13.3$~Jy at $1660$~MHz and $13.1$~Jy at $1713$~MHz, respectively. 
The percentage polarization was taken as $9.6$\% at a polarization angle of $33$\degr. 
In addition, \object{3C138} and \object{3C48} were used as secondary
calibrators. The higher frequency maps have
been convolved with a Gaussian to match the $9\farcm35$ resolution of the \lamA EMLS. 
The rms--noise measured were $15$~mK (\mgls~MHz), $16$~mK ($1660$~MHz) and $19$~mK ($1713$~MHz)
for total intensities and $8$~mK, $9$~mK, and $10$~mK for Stokes~$U$ and $Q$,
respectively. The same conversion factor of $T_\mathrm{B}/S=2.12 \pm 0.02$~mK/Jy at \lamA 
(Uyan{\i}ker et al. \cite{buI}) applies for all frequencies.

In order to remove temperature gradients due to varying ground and
atmospheric radiation from the maps, a
linear baseline was subtracted from each subscan.  This procedure also
affects real sky emission of large spatial
extent, this emission is usually recovered by absolutely calibrated data. However,
other than at \lamA there exist no absolutely calibrated data at \lamB and we calibrated the
maps in the following way: The
temperature spectral index of the Galactic synchrotron emission $\beta^\mathrm{I}$ ($T_\mathrm{B} \propto \nu^{\beta^\mathrm{I}}$) was
adopted to be $\beta^\mathrm{I} \approx -2.7$ in the Taurus area following the results of Reich \& Reich
(\cite{rei88}) for the frequency range between
$408$~MHz and $1420$~MHz. We assumed the same spectral index also for the large-scale 
(LS) polarized intensities ($\beta^\mathrm{PI}_\mathrm{LS}$) and 
calculated an average offset for the \lamB maps from that of the
\lamA map. Rotation measures ($RM$s) across the Taurus region were determined
by Spoelstra (\cite{spoel72}) using data between 408~MHz and 1411~MHz. The $RM$s are very 
low everywhere varying around
zero and we assume an average $RM_\mathrm{LS}$ of $\rm
0~$~\radms to calculate from the extrapolated large-scale polarized
intensity the offsets for the \lamB $U$ and $Q$ maps, respectively.  Since the frequency differences are not
very large we believe that the
applied extrapolation closely resembles absolutely calibrated \lamB polarization data and we expect
only marginal effects on the small-scale polarized structure analysis we
are interested in.

\section{Results}

\subsection{Polarization maps}

\begin{table*}
\centering
\caption{Observed values for the center of the identified Faraday screens in Field 4 
(see Fig.~\ref{bildchen1}b). 
The observed spectral index of the polarized intensity $\beta^\mathrm{PI}_\mathrm{obs}$ and the 
observed $RM_\mathrm{obs}$ were calculated using $\mgls$~MHz and $1660$~MHz data. Errors in $PI$ 
are $11~$mK ($\mgls$~MHz), $13~$mK ($1660$~MHz), and $14~$mK ($1713$~MHz).}
\begin{tabular}{lllllllllll}
\hline
  &   &   & \multicolumn{2}{c}{$\mgls$~MHz} & \multicolumn{2}{c}{$1660$~MHz} & \multicolumn{2}{c}{$1713$~MHz} & & \\
Object & $l$ & $b$ & $PI$ & \hspace{3mm}$PA$ & $PI$ & \hspace{3mm}$PA$ & $PI$ & \hspace{3mm}$PA$ & $\beta^\mathrm{PI}_\mathrm{obs}$ & $RM_\mathrm{obs}$ \\
  & deg & deg & mK & \hspace{3mm}deg & mK & \hspace{3mm}deg &  mK & \hspace{3mm}deg &  & \radm \\
\hline
a & 170.3 & $-9.0$ & 144 & $-29\pm2$ & 136  & $-25\pm2$  & 132 & $-23\pm2$ & $-0.4\pm 0.9$ & $-5\pm4$ \\
b & 170.45 & $-9.4$ & 207 & $-29\pm1$ & 168  & $-22\pm2$  &  163 & $-21\pm2$ & $-1.3\pm0.7$ & $-9\pm3$ \\
c & 170.25 & $-10.0$ & 79 & $-32\pm3$ & 104  & $-33\pm3$  &  102 & $-30\pm3$ & $1.7\pm1.3$ & $1\pm6$ \\
d & 169.75 & $-9.7$ & 97 & $-17\pm2 $ & 124  & $-28\pm2$  &  118 & $-25\pm2$ & $1.5\pm1.1$ & $15\pm5$ \\
e & 169.3 & $-9.55$ & 115 & $-5\pm2$ & 121  & $-24\pm2$  &  108 & $-23\pm3$ & $0.3\pm1$ & $26\pm5$ \\
f & 171.4 & $-8.95$ & 130 & $-18\pm2$ & 142  & $-12\pm2$  &  147 & $-11\pm2$ & $0.5\pm0.9$ & $-8\pm4$ \\
g & 168.15 & $-8.6$ & 75 & $-20\pm3 $ & 103  &  $-30\pm3 $ &  90 & $-29\pm3 $ & $2.0\pm1.4$ & $14\pm6$ \\
h & 174.53 & $-10.18$ & 125 & $-77\pm2$ & 124  &  $-51\pm2$ &  106 & $-55\pm3$ & $0\pm1$ & $-36\pm4$ \\
i & 169.35 & $-10.65$ & 109 & $-23\pm2$ & 117  &  $-22\pm2$ &  111 & $-20\pm2$ & $0.4\pm1.1$ & $-1\pm5$ \\
\hline
\end{tabular}
\label{tablechen1}
\end{table*}

\begin{figure*}
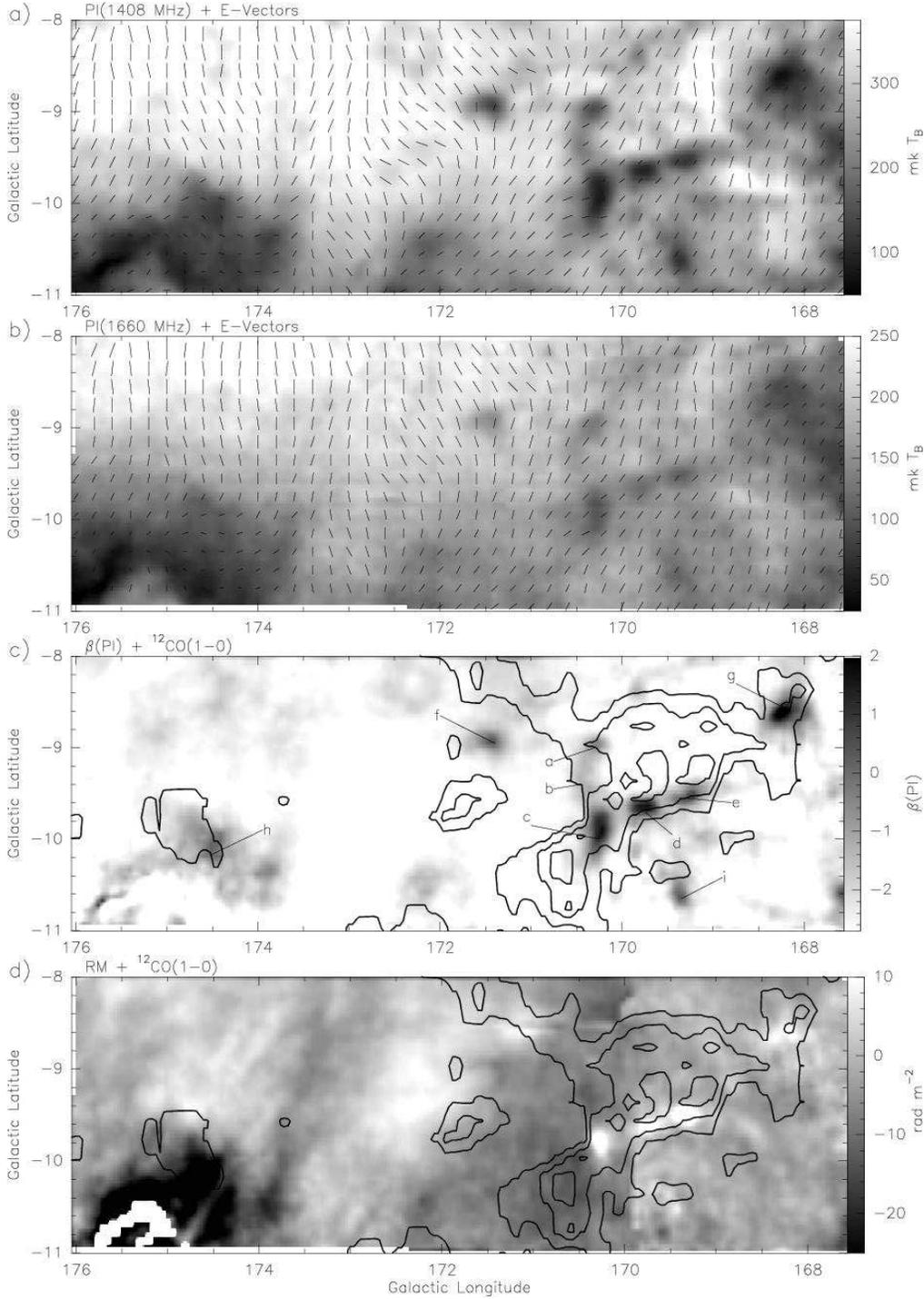

\centering
  \includegraphics[height=13.4cm, angle=270]{0561fig3.jeps}
  \includegraphics[height=13.4cm, angle=270]{0561fig4.jeps}
  \includegraphics[height=13.4cm, angle=270]{0561fig5.jeps}
  \includegraphics[height=13.4cm, angle=270]{0561fig6.jeps}
  \caption{Field 4: Polarized intensity at \mgls~MHz {\bf a)} and $1660$~MHz {\bf b)} with selected polarization E-vectors superimposed, which are
shown for every third pixel (separation $12\arcmin$). The temperature spectral
index of the polarized intensity is shown in {\bf c)}, where in addition the individual objects being modeled 
are indicated.
 The distribution of $RM$ is shown in {\bf d)}. $RM$s are calculated from the polarization data obtained 
at \mgls~MHz and $1660$~MHz. Figures~\ref{bildchen2}c and d show
in addition superimposed contours of integrated \element[][]{CO} emission.} 
\label{bildchen2}
\end{figure*}

The \lamB maps show -- like the \lamA map -- no remarkable structures in total intensities.
All maps show smooth diffuse emission varying mainly with Galactic latitude and a large number of
unrelated extragalactic sources are superimposed. The polarization structures for Fields~1 to 3 
(see Fig.~\ref{bildchen1}b) are rather similar for all wavelengths with marginal differences in structure and 
polarization angles. Neither of these structures show unusual spectral indices of the polarized emission, they are not spatially associated with molecular clouds or any other known objects and originate in the ISM at an unknown distance. For the majority of polarization structures in these fields, the present frequency range and separation seems not adequate for a reliable analysis of the Faraday effects towards these features.

In Field~4 there are a number of small-scale polarization minima, which spatially coincide with the molecular cloud covered in this field. These minima are obviously aligned to the boundary of the cloud. They show rather 
clear differences in intensity and thus in spectral index, as well as in the distribution of polarization angles
between \lamB and \lamAa. For the nine intensity minima we intend to model, we list their
positions, measured polarization data, spectral index and $RM$ in Table~1. 
Spectral index and $RM$ are calculated from the 1408~MHz and 1660~MHz data. No
simple fit based on a $\lambda$ or $\lambda^{2}$ dependence including all data was possible,
reflecting the complex situation of polarized emission transfer through a Faraday screen.
Figures~\ref{bildchen2}a and b show the observed polarized intensities at \mgls~MHz and
$1660$~MHz with selected  E-vectors superimposed. Figure~\ref{bildchen2}c displays the temperature 
spectral index for the polarized intensity calculated from:
\begin{equation} 
\beta^\mathrm{PI}=\frac{\log (PI_{\mathrm{1}}/PI_{\mathrm{2}})}{\log (\nu_{\mathrm{1}}/\nu_{\mathrm{2}})}\ , 
\label{pi_alpha} 
\end{equation}  
with the frequencies $\nu$ and $\nu_{\mathrm{2}}>\nu_{\mathrm{1}}$.
The spectrum of the large-scale emission is close to $\beta^\mathrm{PI}_\mathrm{obs} = -2.7$, which was the assumption
made earlier, while the Faraday screens all show flat or inverted spectra with larger intensities at \lamB
than at \lamAa.
In Fig.~\ref{bildchen2}d we show the distribution of $RM$ which we have
calculated on the basis of the \mgls~MHz and $1660$~MHz observations. While $RM_\mathrm{LS}= 0$~\radms 
was assumed for the large-scale emission we find clear $RM$ deviations in the Faraday screen directions.

\section{Discussion}

\subsection{Additional data for the analysis}
\begin{figure*}
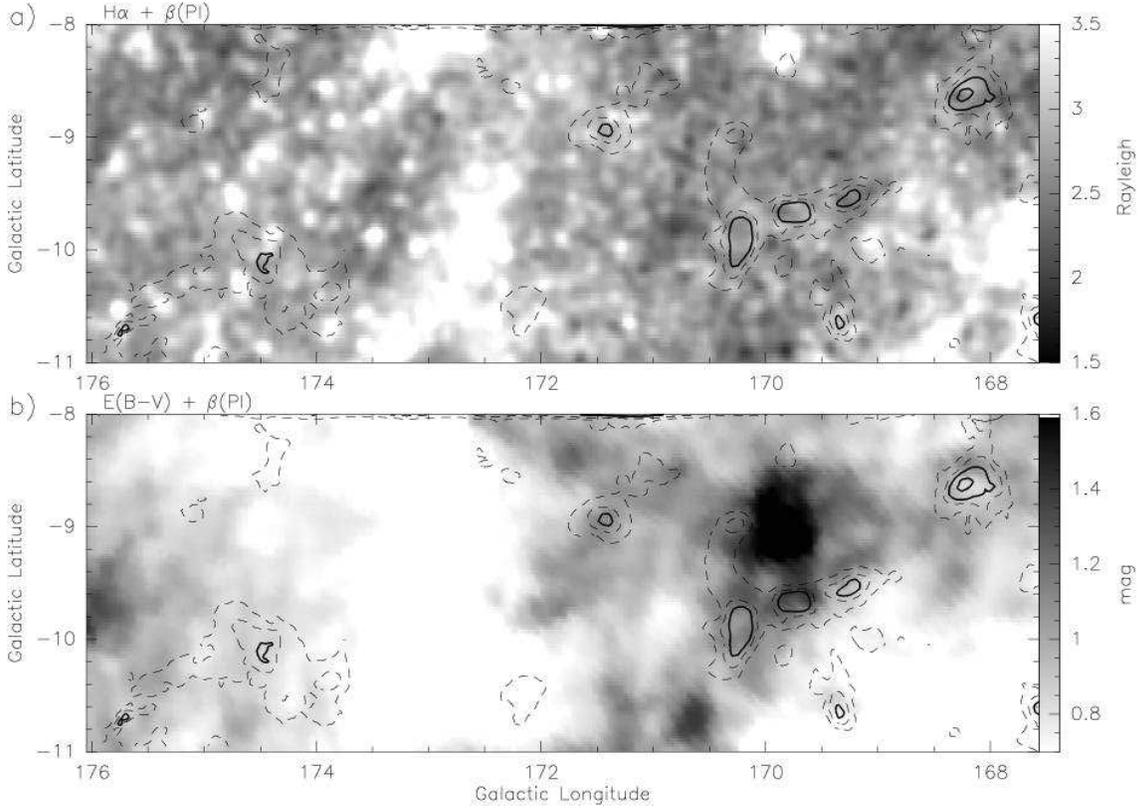

\centering
  \includegraphics[height=15cm, angle=270]{0561fig7.jeps}
  \includegraphics[height=15cm, angle=270]{0561fig8.jeps}
  \caption{{\bf a)} The relevant section from the \char`\"{}full-sky H-alpha map\char`\"{} combined by 
Finkbeiner (\cite{haref}).  {\bf b)} The map
of Schlegel et al. (\cite{schlegel}) showing interstellar reddening $E_\mathrm{B-V}$. 
Both images show selected contours of the temperature spectral index of the polarized 
emission superimposed. Dashed lines indicate negative spectral indices and solid lines positive ones. 
} 
\label{bildchen3}
\end{figure*}

We searched for additional observations of the Taurus region which might be helpful in
the interpretation of the polarization data. $RM$ variations may result from variations
of the thermal electron density, which are traced by H$\alpha$ emission.
In Fig.~\ref{bildchen3}a we show the H$\alpha$ emission taken from the \char`\"{}full-sky 
H-alpha map\char`\"{} which Finkbeiner (\cite{haref}) compiled from northern sky (WHAM and VTSS)
and southern sky (SHASSA) data. 
The angular resolution of Finkbeiner's combined map is $6\arcmin$. No structured H$\alpha$ emission is
visible, which might be correlated with the polarized emission features we have measured. 
The quoted H$\alpha$ sensitivity of $0.52$~Rayleigh is taken as a limit for associated
H$\alpha$ emission.

Correlation of polarized emission structures, however, was found with integrated \element[][]{CO} 
emission and also with interstellar absorption. We display a map 
of the reddening $E_\mathrm{B-V}$ in  Fig.~\ref{bildchen3}b, which is a reprocessed composite of the DIRBE 
and IRAS/ISSA maps at $100~\mu$m by Schlegel et al.
(\cite{schlegel}). The map of Schlegel et al. is believed to trace the dust density more closely than the
IRAS~$100~\mu$m map. Like the map of integrated \element[][]{CO} emission these maps show a remarkable
correlation with the polarization maps. 

\subsection{Faraday screens}

The morphology of the observed polarized emission
and its variation with frequency suggests that strong Faraday rotation
takes place in regions at the surface of molecular clouds.
We use a simple model to describe the effect of these Faraday screens on the
observed distribution of Galactic
polarization. The screen is placed somewhere inside the synchrotron emitting region
and modulates the polarized
emission passing through it. According to the total intensity observations,
where no correlated enhancement is noted, the Faraday screen itself is
assumed not to radiate and the noise level of the total intensity maps
provides an upper limit for its emission. 

Galactic synchrotron emission is generated by cosmic-ray electrons in the
Galactic magnetic field. The position
angle of synchrotron emission, when corrected for Faraday rotation along the
line of sight, is in the
direction of the magnetic field component perpendicular to the line of
sight.  The two emission components in
front and behind the Faraday screen are supposed to be independent with
different intensities and polarization angles, although they might themselves consist 
of multiple layers with different magnetic field orientation and/or some internal 
Faraday dispersion. These emission components are assumed to vary on larger scales 
than the size of the Faraday screens and therefore to have the same properties 
across the area modeled. 
Foreground and background polarization intensities and  polarization angles
are described in the following by 
$PI_\mathrm{fore}$ and $PA_\mathrm{fore}$ or $PI_\mathrm{back}$ and $PA_\mathrm{back}$, respectively.

The Faraday screen may decrease the polarized intensity of the background emission 
by depolarization. In addition it changes the background position angle by
Faraday rotation. Therefore, the observed emission may increase or decrease relative
to its surroundings, depending on the polarization angle orientation of the foreground
and the background emission.

The polarization angle $PA$ of linearly polarized radiation is rotated
by $\Delta PA$ when passing through the
magneto-ionic medium of the Faraday screen and measured at wavelength
$\lambda$. The rotation is given by: 
\begin{equation}
\Delta PA = RM\cdot \lambda^2\ ,
\end{equation}
with $RM$, the rotation measure of the Faraday screen: 
\begin{equation}
RM = 0.81\,B_\parallel\,n_\mathrm{e}\,l\ .
\end{equation}
Here $B_\parallel$ is the magnetic field strength component along the line
of sight measured in $\mu$G,
$n_\mathrm{e}$ is the electron density per cm$^{-3}$ and $l$ is the size of the
Faraday screen along the line of sight measured in pc.

In principle a number of effects are able to lower the observed degree of polarization by depolarization (see
Burn \cite{burn}, Sokoloff et al. \cite{soko}). Due to the relatively small spatial extent of the
Faraday screens discussed here, along with moderate electron densities and magnetic field strengths,
the following mechanisms might potentially contribute: 

\begin{description} 

\item [Beam depolarization:] A variation of the polarization angle across the beam reduces the observed
degree of polarization. A rotation measure gradient within the beam can cause such an angle variation.
$\Delta RM = 17$~\radm$/\mbox{beam}$ at \lamA would reduce the
degree of polarization by $10$\%. We do observe such strong gradients towards the Faraday screens in our field, and we conclude that beam depolarization may not be negligible. Therefore, we include depolarization in our model.

\item [Bandwidth depolarization:] An interstellar rotation measure causes the polarization angle to vary
over the frequency band of the receiver. A $RM$ of $600$~\radms at \lamA is necessary to decrease the
degree of polarization by $10$\%. Such $RM$s are clearly excessive to those observed here
and we conclude that bandwidth depolarization is negligible.

\end{description}

\subsection{The model}
\begin{figure}
  \centering
  \includegraphics[width=6.5cm]{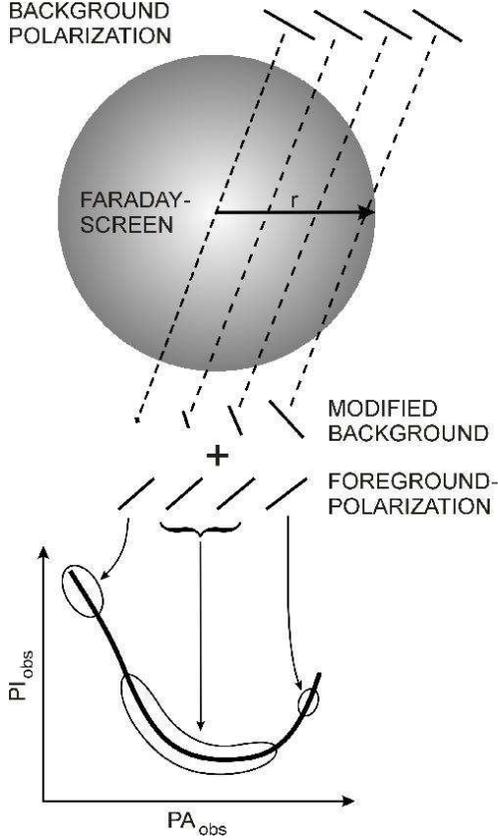}
  \caption{The sketch shows the main components of the model. Polarized emission from behind the Faraday 
screen (background polarization) is systematically rotated and depolarized as a function of path length 
through the screen. The superposition of the modified background and the polarized foreground emission 
leads to a systematic variation of the observed polarized intensity and angle as seen in the $PA$--$PI$ plot.} 
\label{bildchengeo}
\end{figure}

In the following a Faraday screen is assumed to be a uniform sphere with
radius $R$. Elliptical structures were reduced to circulars by normalizing the maximum
radius at each position angle of the ellipse to $1$. The path length for radiation
passing through the screen will therefore vary systematically with position
$r$. With $r$ as the projection of $R$
on the sky, the path length $l$ is given by $l(r)= 2 \sqrt{R^2-r^2}$. We also
assume a constant electron density $n_\mathrm{e}$ and a homogeneous magnetic field ${B}$
inside the sphere with a component $B_\parallel$ along the line of sight. 
Faraday rotation is zero for $r\ge R$. The modified
polarization angle $PA_\mathrm{mod}$ can be written as 
\begin{equation}
PA_\mathrm{mod}(r)=PA_\mathrm{back}+ RM_\mathrm{int}(r)\cdot\lambda^{2}
\end{equation}
with 
\begin{equation}
RM_\mathrm{int}(r) = RM_\mathrm{0}\cdot l(r)\ .
\end{equation}
$RM_\mathrm{0}$ is the maximum intrinsic rotation measure of the Faraday screen at $r=0$.

We assume any intrinsic depolarization $DP_\mathrm{int}$ to increase linearly with the
path length $l(r)$. We express $DP_\mathrm{int}(r)$ by
\begin{equation}
DP_\mathrm{int}(r)=\frac{l(r)}{2R}(1-DP_\mathrm{0})+DP_\mathrm{0}
\end{equation} 
with the maximum intrinsic depolarization $DP_\mathrm{0}$ at $r=0$.
The modified polarized intensity $PI_\mathrm{mod}$ can then be written as:
\begin{equation}
PI_\mathrm{mod}(r)=PI_\mathrm{back}\cdot DP_\mathrm{int}(r)\ .
\end{equation} 
In this notation $DP=1$ stands
for no depolarization and $DP=0$ means complete depolarization
of the polarized background
emission. Depolarization is zero outside the screen ($r\ge R$).

Faraday rotation and depolarization will modify the background polarization
in a systematic way as a function of $r$ (for a schematic representation see Fig.~\ref{bildchengeo}). 
The observed polarization is the superposition of the modified background
and the foreground polarization. In such a case, the polarization properties of the combined emission can be computed by separately adding Stokes parameters $U$ and $Q$. Then $PI$ and $PA$ are determined from these combined Stokes parameters. Given
\begin{equation}
\begin{array}{rl}
U_\mathrm{mod}&=PI_\mathrm{mod}\cdot\sin(2 PA_\mathrm{mod})\\
Q_\mathrm{mod}&=PI_\mathrm{mod}\cdot\cos(2 PA_\mathrm{mod}),
\end{array}
\end{equation}
the measured polarization becomes 
\begin{equation}
\begin{array}{rl}
\label{observable_polarization}
PI_\mathrm{obs}(r)&=\sqrt{\left(U_\mathrm{fore}+U_\mathrm{mod}(r)\right)^{2} + \left(Q_\mathrm{fore}+Q_\mathrm{mod}(r)\right)^{2}}\\
PA_\mathrm{obs}(r)&=\frac{1}{2}\arctan\left(\frac{U_\mathrm{fore}+U_\mathrm{mod}(r)}{Q_\mathrm{fore}+Q_\mathrm{mod}(r)}\right)\ .
\end{array}
\label{vector_components}
\end{equation}

Polarized intensities and position angles in front of and behind
the Faraday screen depend on the observing frequency. In our case we start
from absolute polarization
measurements at \lamA and use a spectral index of $\beta^\mathrm{PI}_\mathrm{LS} = -2.7$
and a $RM_\mathrm{LS} = 0$~\radms
to adjust the relative \lamB data to absolute values as described in Sect.~2. 
The amount of Faraday rotation by the screen is frequency-dependent and
affects $PA_\mathrm{mod}$ and 
therefore $U_\mathrm{mod}$ and $Q_\mathrm{mod}$ accordingly. Since the beam widths of the
observations at the three
frequencies are rather similar we assume depolarization within the Faraday screen 
as well as any depolarization $DP$ not to be frequency-dependent.

Finally the model has four free parameters:
$PI_\mathrm{fore}$ and $PA_\mathrm{fore}$ (or $PI_\mathrm{back}$ and $PA_\mathrm{back}$, respectively) and
the central rotation measure $RM_\mathrm{0}$ and depolarization $DP_\mathrm{0}$ of the Faraday screen.

\begin{figure}
  \resizebox{\hsize}{!}{\includegraphics[bb = 33 45 762 523, clip]{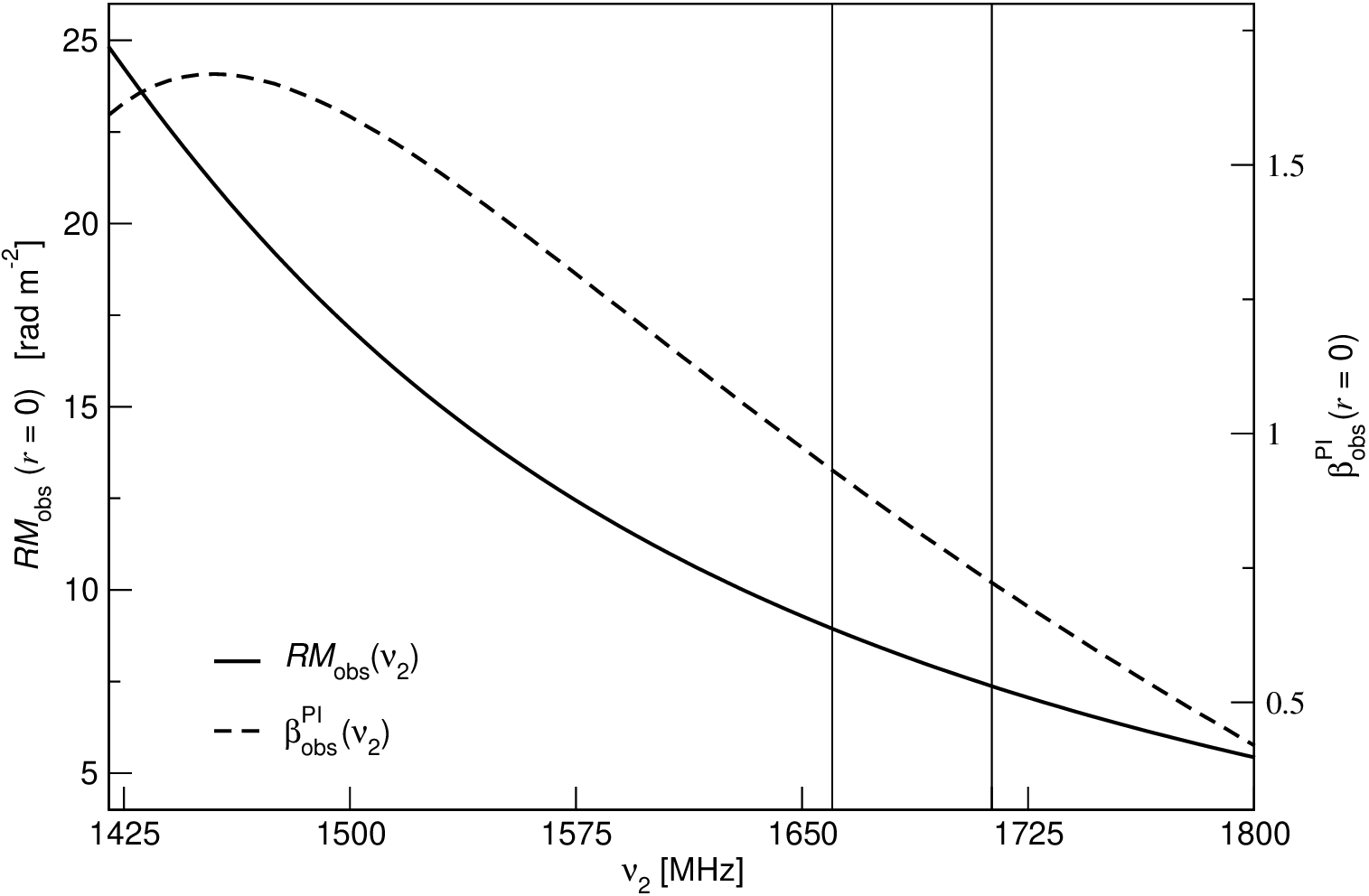}}
  \caption{The plot shows two of the modeled observables for object~d at frequencies between $1410$~MHz and 
$1800$~MHz. The observable rotation measure $RM_\mathrm{obs}$ (solid), as well as the observable 
spectral index $\beta^\mathrm{PI}_\mathrm{obs}$ (dashed) are shown as a function of
the second observing frequency for $r=0$ relative to \mgls~MHz. 
Vertical lines mark the two \lamB frequencies ($1660$~MHz and 
$1713$~MHz) used in this paper.} 
\label{bildchen5}
\end{figure}

\subsection{Notes on the model}  

\begin{enumerate}

\item
The $PA$--$PI$ diagram contains all observed pixels in the field of the Faraday screen as outlined in Fig.~5.
This also holds for the resulting $PA$--$PI$ relation. 

\item
Towards a Faraday screen the observed $RM_\mathrm{obs}$ differs from its intrinsic $RM_\mathrm{int}$ depending
on the amount of foreground polarization adding to the rotated background polarization.
 $RM_\mathrm{obs}$ depends on the two frequencies ($\nu_\mathrm{1},\nu_\mathrm{2}$) used for 
its determination as illustrated in Fig.~\ref{bildchen5}. This also implies no $\lambda^{2}$ dependence
of the observed polarization angles in the direction of Faraday screens. 
Figure~\ref{bildchen5} shows for object~d how $RM_\mathrm{obs}$ and spectral index of $PI$ depend on the second 
observing frequency $\nu_\mathrm{2}$, while $\nu_\mathrm{1}$ is always \mgls~MHz. 

\end{enumerate}

\subsection{Application of the model}

\begin{figure*}
\centering
  \includegraphics[width=14cm]{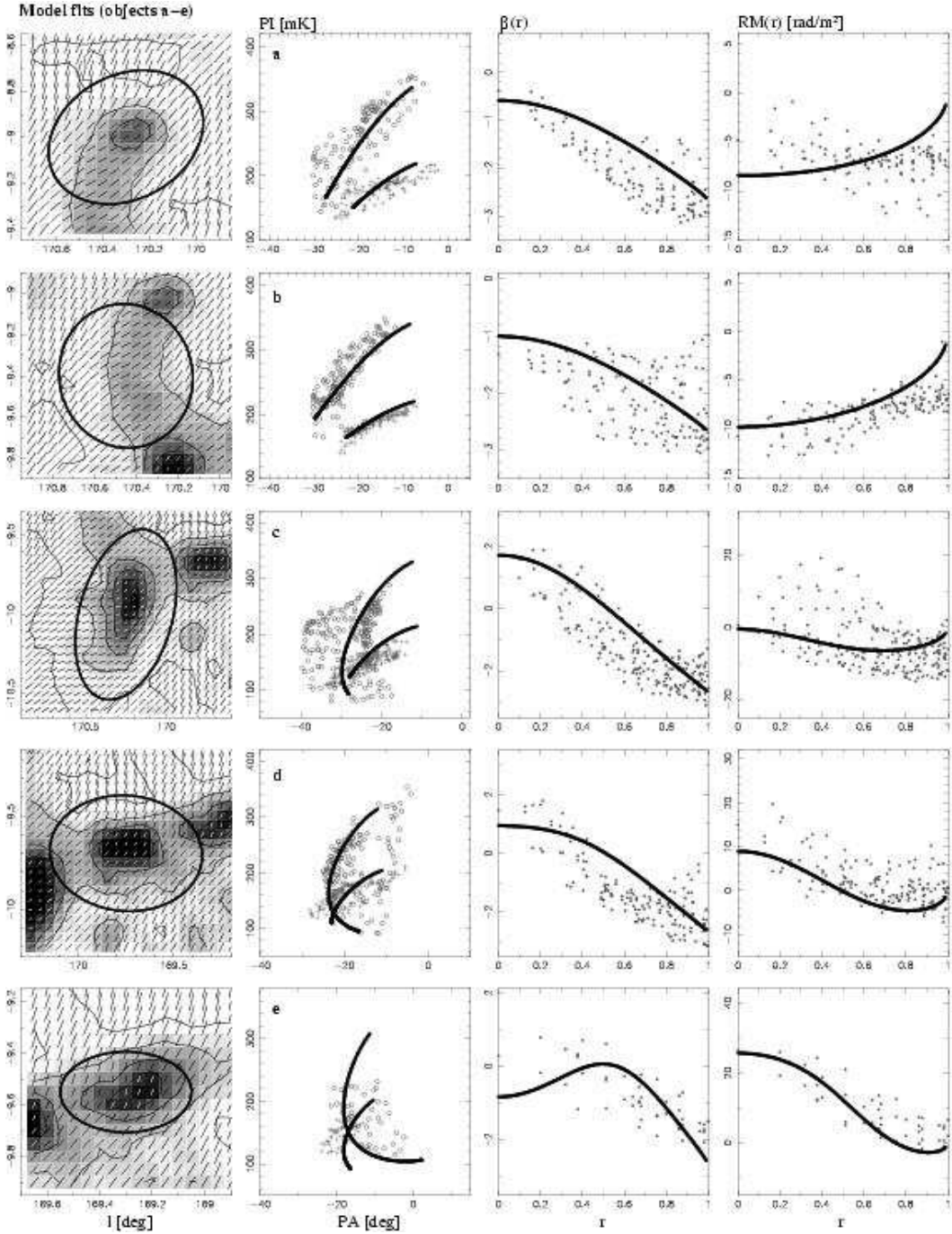}
  \caption{From left to right are shown: The map of $\beta^\mathrm{PI}_\mathrm{obs}$ in Galactic coordinates 
with the ellipse marking the size of the Faraday screen, the $PA$--$PI$ plot for \mgls~MHz (circles) and 
$1660$~MHz (crosses), the observed spectral index of the polarized intensity $\beta^\mathrm{PI}_\mathrm{obs}$, 
as well as the observed rotation measure $RM_\mathrm{obs}$ versus radius $r$. Thick lines indicate the modeled values. Each row refers to a Faraday screen listed in Table~\ref{tablechen1} and labeled from a to i.} 
\end{figure*}
\addtocounter{figure}{-1}

\begin{figure*}
\centering
  \includegraphics[width=14cm]{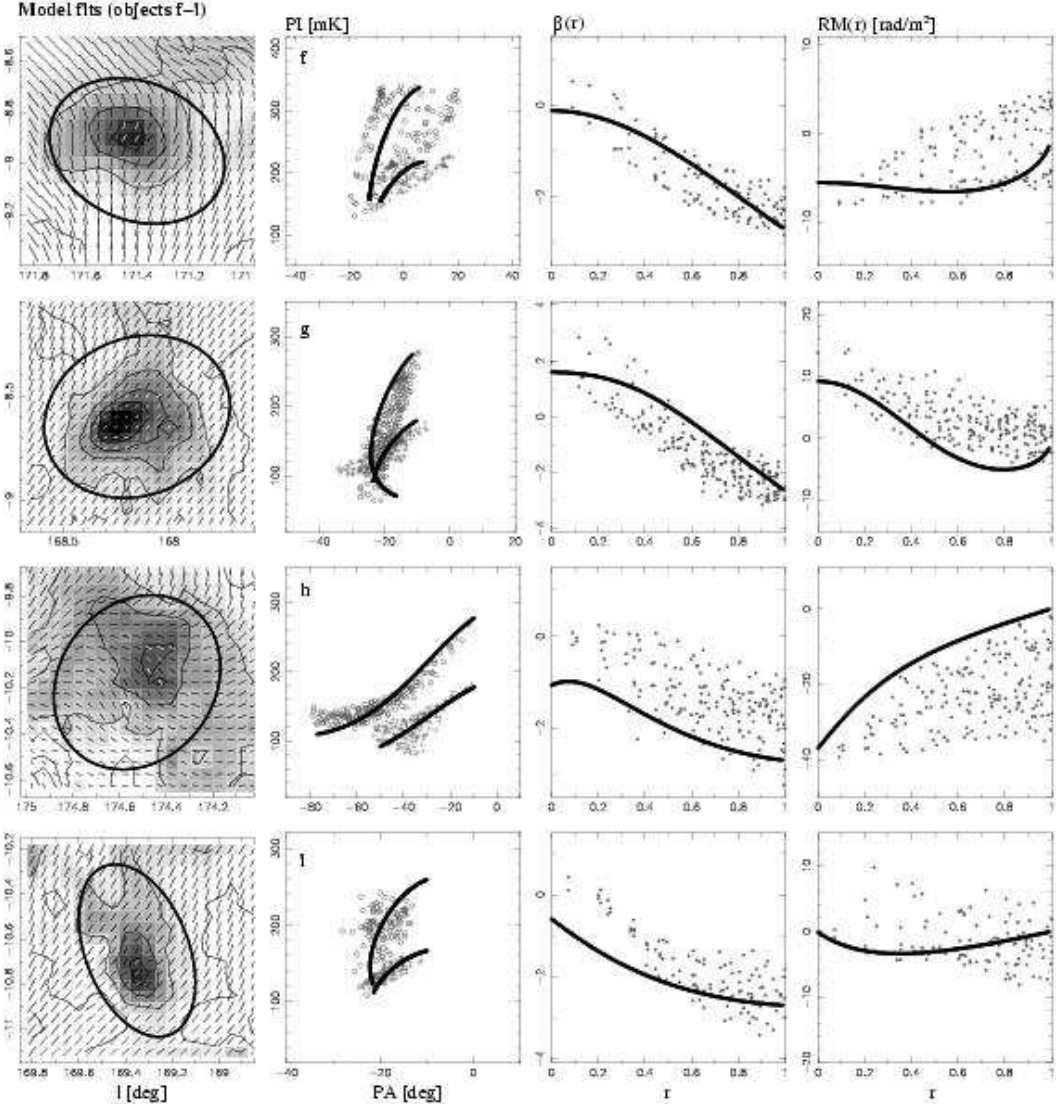}
  \caption{continued} 
\label{bildchen4}
\end{figure*}

\begin{table*}
\centering
\caption{Best--fit parameters of our model fits at \mgls~MHz for the Faraday screens 
listed in Table~\ref{tablechen1}}
\begin{tabular}{llllllll}
\hline
Object & $PI_\mathrm{fore}$ [mK] & $PA_\mathrm{fore}$ [\degr] & $PI_\mathrm{back}$ [mK] &
$PA_\mathrm{back}$ [\degr] & $RM_\mathrm{0}$
[\radm] & $DP_\mathrm{0}$ \\
\hline
a & 186 &  3  & 179 & $-$13 & $-$18 & 1 \\
b & 172 &  0  & 183 & $-$10 & $-$18 & 1 \\
c & 196 & $-$6  & 145 & $-$11 & $-$27.5 & 1 \\
d & 200 & $-$5  & 132 & $-$13 & $-$28 & 0.9\\
e & 224 & $-$2  & 120 & $-$19 & $-$29.5 & 1 \\
f & 202 &  9  & 138 &  11 & $-$25.5 & 1 \\
g & 170 & $-$3  & 124 & $-$13 & $-$27.5 & 0.9\\
h & 82  &  6  & 212 & $-$15 & $-$25 & 0.9 \\
i & 172 & $-$6  &  92 & $-$17 & $-$22 & 1 \\
\hline
\end{tabular}
\label{tablechen2}
\end{table*}

For the nine objects identified as minima in the \lamA polarized intensities
shown in Figs.~\ref{bildchen2}a and b, which also
show unusually high $\beta^\mathrm{PI}_\mathrm{obs}$ as seen from Fig.~\ref{bildchen2}c, we used the following algorithm to
apply our Faraday screen model and find the best-fit parameters:

We start by defining the center of the Faraday screen from the $\beta^\mathrm{PI}_\mathrm{obs}$ map for each object
and estimate radius and ellipticity by eye. This is shown in the left panel of Fig.~\ref{bildchen4}.

For normalized radii $r\ge1$ no Faraday effects are present and the superposition of the foreground and
the background model component must agree with the observed polarization. However, in this case
Faraday screens may partly overlap and influence the offset polarization.

With the boundary condition that the model must correctly reproduce the observed polarization at
$r=1$, four parameters need to be fitted where we allow variations within a reasonable range. 
These limits are:
$50~\mbox{mK}\le PI \le 500$~mK for
the foreground and the background polarized intensity, 
$0\le DP_\mathrm{0} \le 1$ and $-70~\mbox{\radm} \le RM_\mathrm{0} \le 70$~\radm.

Figure~\ref{bildchen4} shows the $\beta^\mathrm{PI}_\mathrm{obs}$ maps, as well as 
$\beta^\mathrm{PI}_\mathrm{obs}$ and $RM_\mathrm{obs}$ versus radius $r$, and 
the model fits for the nine
Faraday screens listed in Table~\ref{tablechen1}. All the pixels of a Faraday screen 
located within the ellipse shown in the left-hand panels of Fig.~\ref{bildchen4} are
displayed in the $PA$--$PI$ diagram.
We calculated the observed spectral index and $RM$ using the \mgls~MHz and $1660$~MHz data. 
The $1713\,$MHz data were used to check the model parameters for consistency.

In order to estimate how sensitive the model is to systematic effects on the 
calibration of the large-scale emission of the \lamB data, 
we re-calibrated the data for different spectral indices and $RM$s for the
large-scale polarized emission ($\beta^\mathrm{PI}_\mathrm{cal} = -2.7 \pm 0.3$ and 
$RM_\mathrm{cal} = 0 \pm 5$~\radm). Subsequent model fitting showed that the best--fit parameters changed 
as follows: $\Delta PI_\mathrm{fore}$, respectively $\Delta PI_\mathrm{back}$ by about $\pm3$~mK, $\Delta 
PA_\mathrm{fore}$ and $\Delta PA_\mathrm{back}$ by about $\pm5$\degr, $RM_\mathrm{0}$ by about 
$\pm5$~\radm. $DP_\mathrm{0}$ remained unchanged.

It is also interesting to compare the results from our model based on the large-scale adjustment of the 
Effelsberg polarization data by the DRAO 1.41~GHz survey presented in this paper with those based on the 
quite limited Dwingeloo data, which were previously used by Wolleben (\cite{woll}) and Wolleben \& Reich (\cite{woll2}). 
For the case of the Taurus region the Dwingeloo-derived
corrections leaves the Effelsberg Stokes $U$ data unchanged and gives a 138~mK offset at 1.4~GHz for the 
Stokes $Q$ map. 
When comparing the Dwingeloo and DRAO polarization intensity scales a difference of 15\% was found in the 
sense that the Dwingeloo intensities are too high (Wolleben et al., in prep.). In addition the DRAO  $U$
and $Q$ maps show some gradients or warping across the field due to their better sampling compared to the Dwingeloo 
data. Running the same model fitting we get the following differences for the average of all clouds
analyzed: $\Delta PI_\mathrm{fore}$, 
respectively $\Delta PI_\mathrm{back}$, by about $-25$~mK and $-12$~mk. $\Delta PA_\mathrm{fore}$ and $\Delta PA_\mathrm{back}$ by about $0$\degr~or $-3$\degr, $RM_\mathrm{0}$ by about 
$-4.3$~\radm~and $DP_\mathrm{0}$ = 1 stays unchanged. 

Although it is essential to take the large-scale polarized intensities for the analysis of Faraday screens
into account, their uncertainties do not change the qualitative results and the quantitative errors are moderate. The model
results for the intrinsic $RM_\mathrm{0}$ of the Faraday screens vary within about 20\%
and allow a conclusive physical interpretation.

\section{Analysis and interpretation}

The parameters for each object we obtained from our model fits are shown in Table~\ref{tablechen2}.
At a frequency of \mgls~MHz, the mean values\footnote{Means were derived by averaging Stokes~$U$ and $Q$ 
separately and calculating $PI$ and $PA$ afterwards. The errors of $PI$ and $PA$ were calculated by the 
error law of Gauss using the standard deviations of $U$ and $Q$ from their means.} 
for the emission component in front of the screen is
$\overline{PI}_\mathrm{fore}=175\pm 40$~mK and
$\overline{PA}_\mathrm{fore}=-1\pm 5$\degr. The background yields
$\overline{PI}_\mathrm{back}=142\pm 37$~mK and
$\overline{PA}_\mathrm{back}=-11\pm 9\degr$ to the observable polarization. The intrinsic rotation 
measures of the Faraday screens average to  $RM_\mathrm{0}=(-25 \pm  4)$~\radm. 
All Faraday screens can be explained with only a small degree of depolarization or none at all.

In the following discussion, the spatial coincidence of polarization minima with the molecular cloud will be interpreted as a strong hint that the Faraday effects, which give rise to the polarization feature, are associated with the molecular cloud. Therefore the distance to the Faraday screens is $140$~pc, the same as it is for the molecular cloud.

Our results show that the background component in most cases is smaller than the foreground 
emission. This is an unexpected result since the foreground emission originates within only
$140$~pc. At a Galactic latitude of about $-10\degr$
the $z$ distance is about $24$~pc and the emission contributing
to the background should originate within several kpc. This may indicate that 
the background layer consists of several superposing emission layers. 
The longer the line-of-sight 
the more likely are changes in the Galactic magnetic field direction and/or Faraday rotation 
within the ISM. This means that depolarization effects increase with the line-of-sight
 and the observed polarized intensity decreases. The fit-results seem to reflects this.

At a distance of about $140$~pc of the molecular clouds the size of the Faraday screens is of the order of
$2$~pc, when spherical symmetry is assumed. Intrinsic $RM_{\mathrm{0}}$ values of up to $\mirm$~\radms require 
an excessive thermal electron density or an excessive regular magnetic field component along the line-of-sight,
when compared to average Galactic values. Since $RM$ is the product of both quantities, further  constrains
are required to derive the appropriate  numbers. We have checked available H$\alpha$ data from the
\char`\"{}full-sky H-alpha map\char`\"{} (Finkbeiner \cite{haref}), which do not show any related
emission ($1\sigma$ detection level of $0.52$~Rayleigh) and constrain $n_\mathrm{e}$ to less than
$\hane$~cm$^{-3}$ for thermal electrons from ionized hydrogen. Upper limits from thermal radio
emission give $n_\mathrm{e} \le 2$~cm$^{-3}$. With an upper limit of $\hane$~cm$^{-3}$ for the 
thermal electron density $n_\mathrm{e}$, a regular magnetic field strength exceeding $\llmf~\mu$G within 
the Faraday screens along the line-of-sight is needed. 

Is the existence of such enhanced localized magnetic field structures in agreement with the observations? 
While magnetic fields directed towards us will not enhance the total intensity synchrotron emission, 
 similar magnetic field enhancements should also exist in
orthogonal directions. We can roughly calculate
the expected fluctuations of the Galactic background emission. Let us assume a magnetic
field with an orthogonal component of $3~\mu$G extending for $1$~kpc along the line-of-sight and giving about
$1$~K brightness temperature, which is a typical value at medium Galactic latitudes at $1.4$~GHz.
 A magnetic field enhancement of $20~\mu$G over $2$~pc, keeping the relativistic electron density
constant, gives about $50$~mK 
for a temperature spectral index of $\beta = -2.7$. Such magnetic bubbles will cause fluctuations of a few
times the confusion limit, which is about $15$~mK~T$_\mathrm{B}$ for the
Effelsberg telescope at $1.4$~GHz. Such fluctuations are quite difficult to
identify in practice, and we conclude that the existence of local magnetic field enhancements 
over size scales of a few parsec is in agreement with the observations. 
Bubbles with field strength exceeding about $30~\mu$G or $5$~pc in size, however, cause total
intensity enhancements of the
order of $150$~mK~T$_\mathrm{B}$. Such structures should be clearly detectable on available radio maps and
therefore are unlikely to exist in large numbers.

\subsection{The origin of Faraday screens}

Although the origin of the Faraday screens discussed here is not clear, we comment on a number of known processes which might be able to cause an enhancement of
n$_\mathrm{e}$ and/or the B-field.

\subsubsection{Ionization in a PDR}

In a photodissociation region (PDR) the physical and chemical properties of the interstellar
medium are dominated by
UV radiation. This is certainly the case for an O,B star ionizing its surroundings, but even the
mean interstellar radiation field can give rise to PDRs. Most of the  atomic and molecular gas in our
Galaxy is exposed to a far-ultraviolet flux and therefore PDRs are expected to be quite common 
(see Hollenbach \& Tielens 1995). 

Models and observations of PDRs show that with increasing extinction $A_\mathrm{V}$
the chemistry changes
from ionized hydrogen at the surface to neutral and finally to molecular hydrogen inside the
cloud. In a similar sequence carbon
exists as \element[][]{C}$^+$ in the outer region and in form of \element[][]{C} and \element[][]{CO} inside.
 According to these models there is a thin layer in the cloud where the ionization of carbon becomes 
the main source for electrons, while the hydrogen-ionizing photons
are already absorbed.

From Fig.~\ref{bildchen3}b we roughly estimate an excess reddening at the positions of the Faraday
screens of $E_\mathrm{B-V}\le0.5\,$mag. Using the usual gas--to--dust relation
$N(\ion{H}{i}+\element[][][][2]{H})/E_\mathrm{B-V} = 5.8\times10^{21}~\mbox{cm}^{-2}\,\mbox{mag}^{-1}$ (Bohlin
et al. \cite{bohlin}), the total neutral hydrogen density amounts to $n\le470~\mbox{cm}^{-3}$
for the size of the Faraday screens of $2$~pc. 

Several authors have derived a C/H ratio of about $10^{-4}$ to $4\times10^{-4}$.
Variations of C/H associated with the physical conditions of the ISM are expected and investigated by
Gnacinski (\cite{gnacinski}). His analysis indicates variations of C/H for different fractional abundances
of molecular hydrogen
$f(\element[][][][2]{H})=2N(\element[][][][2]{H})/(N(\ion{H}{i})+2N\element[][][][2]{H}$. Average values of
C/H$~=3.55\times10^{-4}$ for lines-of-sight with $f(\element[][][][2]{H})<10^{-3}$, and
C/H$~=1.36\times10^{-4}$ for $f(\element[][][][2]{H})>10^{-3}$ were found. For extinctions of 
$A_\mathrm{V}\approx 3.1\,E_{\mathrm{B-V}} \le 1.6\,$mag, 
where we observe the Faraday screens, the PDR model by Hollenbach et al. (\cite{hollenbach1991}) 
gives $f(\element[][][][2]{H})<10^{-3}$. 
In such a case the complete ionization of carbon will enhance the local electron density $n_\mathrm{e}$ 
by about $0.2$~cm$^{-3}$.

This simple estimate shows that a significant part of the thermal electrons in the Faraday screens modeled 
here may be due to carbon ionization in a PDR. In this PDR scenario two extreme cases are possible: 
either no ionization of hydrogen takes place and all electrons are from ionized carbon (total
$n_\mathrm{e}\approx0.2$~cm$^{-3}$) or both elements contribute to a total electron density of $n_\mathrm{e}\le 0.8$~cm$^{-3}$(\ion{H}{ii})$+0.2$~cm$^{-3}$(\element[][]{C}$^+$). With a filling factor of $1$, a magnetic field strength between
$15~\mu$G and $77~\mu$G is required to produce a $RM_{\mathrm{0}}$ of $\mirm$~\radm. Further observations of
H$\alpha$ and \element[][]{C}$^+$ emission could constrain these limits.

\subsubsection{Ionization by cosmic rays}

In clouds with moderate gas densities both photoionization and cosmic-ray ionization could be important.
Theoretical values for the resulting ionization are not well known and spatial variations are expected.
However, referring to McKee (\cite{mckee}) a molecular cloud is primarily ionized by the external UV
radiation field. Only about a tenth of the mass is ionized by cosmic rays. Model calculations by Myers
\& Khersonsky (\cite{myers}) show that for gas densities comparable to the ones adopted here, 
photoionization at the surface of a molecular cloud 
exceeds the ionization fraction from cosmic rays by a factor of $100$. Here an exposure to an UV photon
flux from the mean interstellar radiation field is assumed. Therefore ionization by cosmic rays seems to
be unimportant. 

\subsubsection{Magnetic field enhancement}

A $RM$ enhancement results when the B$_\parallel$ field gets enhanced.
There are a number of possibilities to locally enhance the regular component of the magnetic field in interstellar space. Shock waves from supernova remnants are known to compress the field 
component perpendicular to their velocity direction. Stellar winds or expanding \ion{H}{ii} regions
will also have an effect on the ambient interstellar magnetic field. However, there is no
indication for the presence of a nearby supernova remnant or a massive hot star with a strong stellar
wind in the vicinity of the molecular cloud discussed here. Nevertheless weak shock waves in the 
interstellar medium
without any clear signature may cause some magnetic field compression when they hit a massive molecular 
cloud. Magnetic fields also get
enhanced during the formation process of molecular clouds, although missing excessive synchrotron 
emission limits the magnetic field strength outside the clouds to values not too different from
those in the interstellar space. Zeeman-splitting observations could provide 
the line-of-sight magnetic field strengths or at least an upper limit for the outer regions 
of molecular clouds.

\section{Local synchrotron emissivity}

The local synchrotron emissivity has been previously determined by several
authors. Either the low-frequency emission in front of \ion{H}{ii} regions was
measured or model fits of the Galactic synchrotron emission were made.
The Beuermann et al. (\cite{beuer}) model gives about
$11$~K/kpc for $408$~MHz, which corresponds to about $0.4$~K/kpc at 1408~MHz depending on
the local synchrotron spectral index where we assumed $\beta$ = -2.7.
The Phillipps et al. (\cite{phil}) 408~MHz model of the Galaxy places the Sun near a local
minimum. They report even lower values corresponding to less than $0.14$~K/kpc at 1408~MHz. 
For the case of low-frequency measurements towards optically thick \ion{H}{ii} regions as discussed 
by Fleishman \& Tokarev (\cite{fleishman})
at 10~MHz and Roger et al. (\cite{roger}) at 22~MHz an enhanced local volume synchrotron 
emissivity is indicated. However, this is largely based on the nearest \ion{H}{ii} region $\zeta$-Oph (Sh 27) 
at $\sim$~170~pc distance, which shows excessive foreground synchrotron emission as discussed
by Roger et al. (\cite{roger}). Excluding $\zeta$-Oph, Roger et al. (\cite{roger}) report a mean value of 30~K/pc
at 22~MHz, which corresponds to $0.9$ to $0.6$~K/kpc at 1408~MHz for $\beta$ of -2.5 or -2.6, respectively.

Our mean polarized emission at 1408~MHz in
front of the Taurus molecular cloud is about $\pfe$~K, which requires a synchrotron
emission of at least $\sfe$~K for the maximum possible percentage polarization.
The distance to the Taurus cloud is $140$~pc. Assuming the same distance for the Faraday screens, this value corresponds 
to an emissivity of about $\tse$~K/kpc. Since this is a lower limit set by the 
maximum possible polarization ($\sim 75$\%), the local emissivity is likely significantly higher.
Other than for the \ion{H}{ii} region $\zeta$-Oph (Sh 27), which is at a similar distance,
there is no indication for excessive foreground emission towards the Taurus molecular clouds.
However, the local emissivity may depend on direction. This may be investigated by
measurements of this kind towards molecular clouds in other areas of sky.  
In case the local synchrotron emissivity derived towards the Taurus cloud 
holds for all directions, synchrotron models of the Milky Way (Beuermann et al. \cite{beuer}; 
Phillipps et al. \cite{phil}) may need modifications. Enhanced local 
emissivity has a significant effect when modeling the thick disk properties 
of the Galaxy. The large height of $\sim$5~kpc or more inferred by the Beuermann et al. (\cite{beuer}) and Phillipps et al. (\cite{phil}) models
will be reduced or its emissivity decreases. It should be noted that the modeled Galactic halo 
extends several kpc above the plane. Such halos are rare and place the Milky Way within a small group of 
galaxies showing similarly huge halos (e.g. Dumke \& Krause \cite{dumke}).

\section{Conclusions}

We have used the Effelsberg telescope for observations of the polarized emission in the Taurus region
and added the missing large-scale components by making use of a new DRAO 1.4~GHz polarization survey.   
We observe minima in polarized intensity along the boundary 
of a molecular cloud complex located at a distance of $140$~pc. 
Excessive $RM$s were measured based on 1408, 1660 and 1713~MHz data. 
The polarization angles do not follow a $\lambda^{2}$ dependence, which indicates the
presence of components with different physical properties seen in superposition along the line-of-sight. 
The observed systematic variation of polarization angles with polarized intensity across the Faraday
screens were fitted by a model assuming a constant large-scale foreground
and also background emission. The large-scale components at \lamB were scaled from absolutely calibrated
\lamA data with a spectral index of $\beta^\mathrm{PI}_\mathrm{LS} = -2.7$
and $RM_\mathrm{LS} = 0$~\radm. We calculate physical parameters for the magneto--ionic
medium at the surface of the molecular clouds, which differ largely
from average interstellar values. Since there is no observational evidence for enhanced thermal 
electron density
from available data, an enhanced $B$-field is needed to account for the large $RM_{\mathrm{0}}$ observed. 
Towards Faraday screens our model predicts significant variations of $RM_\mathrm{obs}$ for
different frequency intervals (see Fig.~\ref{bildchen5}). Recently published $RM$ maps 
such as those of Haverkorn et al. (\cite{haver2}), Gaensler et al. (\cite{gaens}) and Uyan{\i}ker et al. (\cite{buIII}) 
are based on fits assuming a $\lambda^{2}$ dependence of multi-channel data. The fit is often imperfect or 
impossible. This may be due to limitations in the measurement accuracy. 
However, the superposition of various components of the magneto--ionic interstellar medium results in deviations 
from a $\lambda^{2}$ dependence and this should be taken into account when
quoting $RM$ values.

From the model fits presented we find quite high
values for the foreground synchrotron emissivity in the direction of the Taurus molecular clouds. 
If this emissivity represents the omnidirectional local emissivity, it is
several times larger than previously assumed. This will have consequences for the extent of
 the large-scale-halo or the thick disk emission of the Galaxy, which will shrink in size or 
is weaker than previously assumed. 

\begin{acknowledgements}
We like to thank Ernst F\"urst, Tom Landecker and Richard Wielebinski for critical reading of the manuscript
and discussions. We also thank the anonymous referee for useful comments and improvements of the manuscript.
\end{acknowledgements}

\end{document}